\documentclass[12pt]{article}
\usepackage[all]{xy}

\usepackage{epic}
\usepackage{amsmath}[1996/11/01]
\usepackage{amssymb,amsthm,amsfonts,latexsym,epsfig,graphics}
 \def\beql#1#2\eeql{\begin{equation}\label{#1}#2\end{equation}}

\advance \textheight by -1 \topmargin
\advance \textheight by -1 \headheight
\advance \textheight by -1 \headsep
\advance \textheight by -2 \footskip
\vsize = \textheight
\hsize = \textwidth
\DeclareMathOperator{\wt}{wt}

\DeclareMathOperator{\Aut}{Aut}
\DeclareMathOperator{\soc}{soc}

\newtheorem{theorem}{Theorem}[section]

\newcommand{\bew}{\noindent\underline{Proof.}\ }

\newtheorem{lemma}[theorem]{Lemma}

\newtheorem{folg}[theorem]{Corollary}

\newcommand{\F}{{\mathbb{F}}}

\newcommand{\eb}{\phantom{zzz}\hfill{$\square $}\smallskip}

\renewcommand{\em}{\sf}

\begin{document}
\begin{center}
{\Large {\bf An extremal [72,36,16] binary code has no automorphism
group containing $Z_2\times Z_4$, $Q_8$, or $Z_{10}$.}} \\
\vspace{1.5\baselineskip}
{\em Gabriele Nebe\footnote{
Lehrstuhl D f\"ur Mathematik, RWTH Aachen University,
52056 Aachen, Germany, 
 nebe@math.rwth-aachen.de}}
\end{center}

{\sc Abstract.}
{\small
Let $C$ be an extremal self-dual binary code of length 72
and $g\in \Aut(C) $ be an automorphism of order 2.
We show that $C$ is a free $\F_2\langle g \rangle $ module and 
use this to exclude certain subgroups of order 8 of $\Aut (C)$.
We also show that $\Aut(C)$ does not contain an element of order 10.
Combining these results with the ones obtained in earlier papers we find that 
the order of $\Aut(C)$ is either 5 or divides 24. 
If 8 divides the order of $\Aut(C)$ then its Sylow 2-subgroup is either 
 $D_8$ or $Z_2\times Z_2\times Z_2$.
}

\section{Introduction.}

Let $C=C^{\perp }\leq \F_2^n$ be a binary self-dual code. 
Then the invariance properties of the weight enumerator of $C$ and
its shadow may be used to obtain an upper bound for the minimum distance 
$$d(C):= \min \{ \wt(c) \mid 0 \neq c \in C \} \mbox{, where }
\wt(c) := | \{ i \mid c_i \neq 0 \} | ;$$ 
$d(C) \leq 4 \lfloor \frac{n}{24} \rfloor + 4 $
unless $n\equiv _{24} 22$ where the bound is $4 \lfloor \frac{n}{24} \rfloor + 6 $
(see \cite{Rains}).
Codes achieving equality are called extremal. 
Of particular interest are extremal codes of length $24k$. 
For $n=24$ and $n=48$ there are unique extremal codes \cite{QRunique}, 
both are extended quadratic residue codes. 
For $n=72$ the extended quadratic residue code fails to be extremal and 
no extremal code of length 72 is known. 
One frequently used method to search for an extremal code is to investigate
codes which are invariant under a certain subgroup of the symmetric group $S_n$.
For $n=72$ it has been shown in a series of papers that the 
automorphism group 
$$\Aut(C)  := \{ \sigma \in S_n \mid \sigma (C) = C \} $$ of an extremal code is either
$Z_5$ or $Z_{10}$ or its order divides $24$
(for more details and references see \cite{FN}). 
This paper introduces a new method and excludes the cases 
that $\Aut(C)$ contains a quaternion group of order 8, the 
cyclic group $Z_{10}$, or the group $Z_4\times Z_2$.
It also provides a new proof that $\Aut(C)$ does not contain an
element of order 8.

\section{Indecomposable modules for cyclic groups.}

The main result from modular representation theory that we use  in this note
is the classification of indecomposable $\F G$-modules  for cyclic $p$-groups $G$
over a field $\F$ of characteristic p.
By the theorem of Krull Schmidt any $\F G$-module 
is up to isomorphism a unique direct sum of indecomposable modules.

\begin{theorem} \label{Alperin} (see for instance \cite[pp 24,25]{Alperin})
Let $G = \langle g \rangle $ be a cyclic group of order $q:=p^a$ and 
$\F $ a field of characteristic $p$.
Then the group ring $\F G $ is isomorphic to $R:=\F[X]/(X^q)$
via $g\mapsto X+1$.
This ring $R$ is a uniserial ring with ideals $(X^i) $, $0\leq i \leq q$.
All indecomposable $R$-modules are the factor modules
$V_i := R/(X^i) $ for $1\leq i \leq q$.
The module $V_1 \cong S$ is the simple $\F G$-module and $V_q = R $ is the
free module of rank 1.  
\end{theorem}

\begin{folg}\label{restrict}
Let $G$ be a cyclic $p$-group and
$M$ an $\F_p G$-module.
 Then $M$ is projective if and only if $M$ is free
if and only if the restriction of $M$ 
to the subgroup of order $p$ is free.
\end{folg}

There are not many groups where such a strong criterion holds. 
One other group is the quaternion group $Q_8$ of order 8. 
More precisely \cite{Carlson} shows the following.

\begin{lemma} \label{Q8} 
Let $G=Q_8$ be the quaternion group of order 8 and $Z:=Z(G)$ the
center of $G$, of order $2$. 
Let $M$ be an $\F_2 G$-module. 
Then $M$ is free as an $\F_2G$-module if and only if 
$M$ is free as an $\F_2 Z$-module.
\end{lemma}

\section{The main result.}

Throughout this section
let $C = C^{\perp } \leq \F_2^{72}$ be a self-dual code with minimum
distance $d(C) =16$ 
and $g\in \Aut(C) $  be an automorphism of order 2. 
By \cite{Bouknofix} the permutation $g$ has no fixed points, so 
we may assume that 
$g = (1,2)(3,4)\ldots (71,72) $. 
Let 
$$C(g):= \{ c\in C \mid g(c) = c \} = \{ c = (c_1,c_1,c_2,c_2,\ldots , c_{36},c_{36}) \in C \} $$ be the fixed code of $g$. Define two mappings 
$$\begin{array}{ll}
\pi : C(g) \to \F_2^{36} , &  (c_1,c_1,c_2,c_2,\ldots , c_{36},c_{36}) 
\mapsto (c_1,\ldots ,c_{36}) \\
\Phi : C \to \F_2^{36} , & (c_1,\ldots , c_{72}) \mapsto (c_1+c_2,c_3+c_4,\ldots , c_{71}+c_{72}) .\end{array} $$
Then $\Phi (C) \subseteq \pi (C(g)) = \Phi (C)^{\perp } $ (see \cite{Bouknofixa}).

\begin{theorem}\label{main}
Let $g$ be an automorphism of order $2$ of an extremal self-dual
code $C$ of length $72$.
Then $C$ is a free $\F_2\langle g \rangle $-module and $\pi (C(g)) = \Phi (C)$ is
a self-dual $[36,18,8]$ binary code.
\end{theorem}

\bew
We consider $C$ as a module for the group algebra $R:=\F_2 \langle g \rangle $.
By Theorem \ref{Alperin} 
the ring $R$ has up to isomorphism two indecomposable modules,
the free module $R$ and the simple module $S \cong \F_2$.
The module $C$ has $\F_2$-dimension 36 and hence is of the form
$C \cong R^a \oplus S^{36-2a} $. Clearly the
fixed code $C(g)$ is the socle of this module; 
$C(g)= \soc (C) \cong S^{a} \oplus S^{36-2a} $ 
 of dimension $36-a$.
So $\pi (C(g)) \leq \F_2^{36}$ has dimension $36-a$ and
the minimum distance is $d(\pi (C(g))) \geq \frac{d(C)}{2} = 8 $.
By the above $\pi (C(g))  = \Phi (C)^{\perp } \geq \Phi (C) $
is the dual of a self-orthogonal code and hence contains some self-dual
code $D=D^{\perp }\leq \F_2^{36}$ of minimum distance $\geq 8$.
By \cite{Gaborit} there are 41 such codes.
With Magma \cite{MAGMA} one checks that 
no proper overcode of these 41 codes has minimum distance $\geq 8$, so
$\dim (C(g)) = 18$ and hence $a=18$.
Therefore $C\cong R^{18}$ is a free $\F_2 \langle g \rangle $-module and 
$\pi (C(g))$ is one of these 41 extremal self-dual codes.
\eb

The first corollary
 also follows from the Sloane-Thompson theorem (see \cite{ST}, \cite{NebeGuenther})
since any extremal code of length $24k$ is doubly even
(see \cite{Rains}):

\begin{folg}
Let $C = C^{\perp }$ be an extremal code of length $72$. 
Then $\Aut (C) $ does not contain an element of order 8.
\end{folg}

\bew
Assume that there is some $\sigma \in \Aut(C)$ of order 8.
Since $C$ is free as a module over $\F_2\langle \sigma ^4 \rangle $
by Theorem \ref{main},
Corollary \ref{restrict} says that 
$C$ is a free $\F_2 \langle \sigma \rangle $-module;
so $C\cong \F_2 \langle \sigma \rangle^a$ with
 $8a = \frac{72}{2} = \dim(C) $. Thus $a=9/2$, a contradiction.
\eb

\begin{folg}
Let $C = C^{\perp }$ be an extremal binary code of length $72$. 
Then $\Aut (C) $ does not contain a quaternion group of order 8.
\end{folg}

\bew
Assume that there is some subgroup $G\leq \Aut(C)$ such that 
$G$ is isomorphic to the quaternion group of order 8.
Let $Z := Z(G)$ be the center of $G$. This is a group of order 2
and so by Theorem \ref{main} the code $C$ is a free $\F_2Z$-module.
By Lemma \ref{Q8} this implies that $C$ is also a free $\F_2G$ module
of rank $\dim(C)/8 = 9/2$, which is absurd.
\eb

\begin{folg}
Let $C = C^{\perp }$ be an extremal  binary code of length $72$. 
Then $\Aut (C) $ does not contain a subgroup $Z_4\times Z_2$.
\end{folg}

\bew
Assume that $U\cong Z_4\times Z_2 = \langle h , g \rangle $ is a subgroup of $\Aut(C)$ such that 
$g^2=h^4=1$.
Since any element of order 2 in $U$ acts fixed point freely on $\{ 1,\ldots ,72 \}$,
the group $U$ acts freely on this set and $\langle h \rangle \cong Z_4 $ acts freely on 
the set of $\langle g \rangle $-orbits. 
Therefore $h$ acts as a permutation with nine 4-cycles 
 on the the fixed code $C(g)$.  A Magma computation 
shows that none of the 41 self-dual 
$[36,18,8]$ codes 
from \cite{Gaborit}  has such an automorphism.
\eb

The following corollary summarizes these three results.

\begin{folg}
Let $C = C^{\perp }$ be an extremal  binary code of length $72$. 
If $8$ divides $|\Aut(C) |$ then the Sylow $2$-subgroup of $\Aut(C)$ is 
either $Z_2\times Z_2\times Z_2$ or $D_8$.
\end{folg}

\begin{folg}
Let $C = C^{\perp }$ be an extremal  binary code of length $72$. 
Then $\Aut (C) $ does not contain an element of order $10$.
\end{folg}

\bew
By \cite{CP} any element of order 5 in $\Aut (C)$ has fourteen 5-cycles and 
two fixed points. 
If there is some $\sigma \in \Aut(C)$ with order 10, then $\sigma ^2$ acts on the 
fixed code $C(\sigma ^5)$ of the element of order 2 as a permutation with 
seven 5-cycles and one fixed point. A Magma computation 
shows that none of the 41 self-dual 
$[36,18,8]$ codes 
from \cite{Gaborit}  has such an automorphism 
of order 5. This is shown independently in \cite{YY}.
\eb

{\bf Acknowledgement}. All computations have been done in Magma \cite{MAGMA}.

\end{document}